# Fringe field control of one-dimensional room temperature quantum transport in site controlled AlGaN/GaN lateral nanowires

Akhil K. S[†], Dolar Khachariya[†], Mudassar Meer, Swaroop Ganguly and Dipankar Saha[*]

Applied Quantum Mechanics Laboratory, Centre of Excellence in Nanoelectronics, Indian Institute of Technology Bombay, Powai, Mumbai – 400076, India

*Email: dipankarsaha@iitb.ac.in, Phone: +91-22-25767443

[†]Equal contributing authors.

Abstract:

We have demonstrated effective fringe field control of one-dimensional electron gas (1-DEG) in AlGaN/GaN lateral nanowires. The nanowires are site controlled and formed by a combination of dry and anisotropic wet etching. The nanowire dimensions are well controlled and can have a very high length/width aspect ratio of 10 µm/5 nm or larger. The transport is controlled by a fringe gate and shows room temperature quantum transport where gradual filling of 1-D subbands gets manifested as oscillations in the transconductance. The fringe gate threshold voltage for depletion of one-dimensional electron gas is found to increase with increasing drain voltage indicating efficient control of 1-DEG. The transport characteristics and fringe field operation are explained by taking into account quantum capacitance in addition to the conventional geometric capacitance. The effect of nanowire width and fringe gate position is also discussed.

GaN and its alloys have received a lot of attention for its immense potential for various applications both in electronic and optoelectronic devices.[1-5] The epitaxial growth of high quality GaN, InN, AlN and their alloys have led to the development of various quantum mechanical devices with improved functionality and enhanced performances.[6-11] The nanostructures lend themselves to many advantages through quantum confinement and GaN based quantum wires have found ubiquitous usage in LEDs, lasers and transistors.[12-15] The nanowire transistors are amenable for gate-all around structures for high performance. They are also frequently used to study quantum transport in reduced dimensional structures.[16-18] Most of the nanowires on GaN are grown vertically and dislodged before drop casting on to a substrate to form lateral nanowires.[16,19,20] Nanofabrication techniques are then used to lithographically isolate individual nanowires. The dimension and position of the nanowires are difficult to control in this method. We have demonstrated a new method where site controlled AlGaN/GaN lateral nanowires can be formed from a quantum-well structure by using a combination of dry and anisotropic wet chemical etching processes. The aspect ratio of the nanowire can be very high. The nanowire widths can be made very small (< 5 nm) and limited by the detection limit of the inspection tool; while they can be very long (> 10 μm). We have shown that fringe electric fields arising out of a lithographically defined non-contacting gate can control the 1-DEG channel transport in these nanowires. The fringe electric field from a non-contacting terminal, which is usually considered to have detrimental effects, is demonstrated to have the potential to work as the sole gate for a transistor. The large lateral confinement in the nanowires lead to an increased electron 1-D intersubband separation resulting in room temperature quantum transport in these devices, which manifest as oscillations in the transfer characteristics. The 1-D transport characteristics is explained by taking into account quantum capacitance in addition to the geometric capacitance.[21,22]

Figure 1(a) shows a schematic of the heterostructure used in this work. A 2-DEG exists at the AlGaN/GaN interface, which is converted to 1-DEG after nanowire formation. Figure 1(b) shows a schematic of the nanowire and the fringe gate fabricated from the heterostructure. Two ohmic contacts are made to the end of the nanowires, which act as source (S) and drain (D) regions. The non-contacting fringe gate (FG) is delineated at perpendicular to the nanowire. Figure 1(c) shows the equilibrium band diagram as obtained by solving 2-D Schrodinger and Poisson equations self-consistently. The additional lateral confinement pushes the bound states further up with respect to the conduction band edge ($E_C$) and increases the separation of 1-D intersubbands. As the fringe gate voltage ($V_{FG}$) becomes negative the number of available bound states below Fermi energy ($E_F$) decreases. The electron density encounters quantum jumps whenever a bound state approximately pass the Fermi level ($E_F$) and shows plateaus between two successive sub-bands crossing the $E_F$. This happens due to the inherent nature of the 1-D density of states which has a long but narrow tail for higher energy levels. Figure 1(d) shows the typical fringe electric fields lines of force around the nanowire for negative gate voltage. The electric field controls the 1-DEG channel from the top surface and two side-walls (separated by ~10 nm). The fringe gate action on a nanowire is equivalent to a tri-gate transistor and can control the 1-DEG channel conductance efficiently.

The nanowires are fabricated by a combination of dry and anisotropic wet chemical etching processes on the heterostructure shown in Fig. 1(a). A scanning electron microscopic (SEM) image of a typical nanowire is shown in the inset of Fig. 1(e). A cross sectional SEM image of the nanowire is shown in Fig. 1(e). The fabrication steps are described in the supplementary document. It is trapezoidal in shape and the angle between the vertically exposed planes is found be 119° which matches closely with the expected angle of 116° between the side-planes of Miller indices $(10\bar{1}3)$ and $(\bar{1}013)$ (see supplementary document).[23] Figure 1(f) shows a colored SEM image of a fringe-field nanowire device. It may be noted that experimental results from a single fringe gate is reported here. The two gates at different separations from the nanowire allow study of transport for various gate to channel couplings. It is possible to fabricate an array of such nanowire transistors by starting with a pattern having periodic rectangular regions defined at desired sites as shown in Fig. 1(g). The photoluminescence (PL) measurement over a large area of the sample after chemical wet-etching is shown in Fig. 1(h). The presence of a dominant peak at 331 nm corresponds to the AlGaN .The AlGaN peak indicates the presence of the channel in spite of the blanket wet-etching. The 1-DEG is formed at the AlGaN/GaN interface. The etch-time determines the lateral dimension of the nanowire.

Figures 2(a) and (b) show the measured drain to source current ($I_{DS}$) as a function of the fringe gate voltage ($V_{FG}$) for various values of drain voltages ($V_{DS}$) in linear and log scale, respectively. The transconductance $g_m$ (= $\partial I_{DS}/\partial V_{GS}$) is shown in Fig. 2(c). The $I_{DS}$ is dominated by rapid increases interleaved by plateaus, which manifest as oscillations in $g_m$. This corresponds to the quantum jump in electron density as the Fermi level ($E_F$) *approximately* crosses a bound state and the almost constant electron density when $E_F$ is between the two sub-bands. The number of bound states and their energy levels are a function of both $V_{DS}$ and $V_{FG}$. However, it is primarily determined by $V_{FG}$ for small values of $V_{DS}$ when the transistor is in the linear region. Hence, a knowledge of the fringe field coupling to the nanowire and the separation of plateaus in the $I_{DS}$- $V_{FG}$ characteristics (for small value of $V_{DS}$) allow to determine the 1-D sub-bands. The $g_m$ for small values of $V_{DS}$ is given by (see supplementary dpcument):

$$g_m = \mu \frac{V_{DS}}{L}\left[\frac{e^2\, \partial n_s/\partial E_F}{1 + e^2 C_{FG}^{-1}\, \partial n_s/\partial E_F}\right] = \mu \frac{V_{DS}}{L} C_T \qquad (1)$$

where $\mu$ is the mobility, L is the nanowire length, e is the electronic charge, and $C_{FG}$ is the geometric fringe capacitance. It may be noted that the term in the bracket (total capacitance, $C_T^{-1} = C_{FG}^{-1} + C_Q^{-1}$) corresponds to a series combination of $C_{FG}$ and the quantum capacitance ($C_Q = [e^2\partial n_s/\partial E_F]$). $C_Q$ undergoes oscillations as a function of $V_{FG}$, which in turn gives rise to oscillations in $g_m$ [$1/g_m = (\mu V_{DS}/L)*(1/C_{FG}^{-1} + 1/C_Q^{-1})$]. The position of the maxima and minima can be determined theoretically (see supplementary document).

The theoretical position of the peaks does not correspond exactly where the $E_F$ crosses the lower energy bound state ($E_1$). This can be explained by noting $C_q$ as a function of $E_F$ shown in Fig. 3(a). The inter-subband separation is assumed to be large (100 meV) here. It may be noted that $C_q$ shows peaks for $E_F - E_1 = 1.27\, k_B T$, where $k_B$ is the Boltzmann constant and T is the temperature; when ($E_1 - E_2$) is very large. The apparent deviation from the intuition is due to the fact that $n_s$ experiences a sudden jump only when Fermi-Dirac function is shifted towards higher energy level in comparison to $E_1$ and $E_2$. However, it is to be further noted that the relative magnitude of the peaks for $C_q$ due to $E_1$ and $E_2$ is also dependent on the inter-subband separation, $\Delta E = E_1 - E_2$ in a generalized scenario. The distinct peaks may disappear and converge to a single peak when $\Delta E$ is small. The estimated inter-subband separation is found to be 0.22 eV for the ~10 nm nanowire transistor by matching experimental data with theory as shown in Fig. 4(b). It is assumed that $E_F$ ($E_1 < E_F < E_2$) varies linearly with $V_{FG}$ in this case. The energy band-diagrams for various positions of $E_F$ with respect to $E_1$ and $E_2$ are shown in Figs. 3(c)-(e).

The $I_{DS}$ for larger values of $V_{DS}$ is complicated by the fact that the channel potential is significantly affected by the drain potential and the number of subbands carrying the current may vary along the channel. The number of bound states is more near the source end and it decreases towards the drain end. Hence, $I_{DS}$ with larger values of $V_{DS}$ will show oscillations when $V_{FG}$ is large enough that it brings the transistor back in the linear regime. This is also experimentally observed in Fig. 2(b) where the originating voltage of plateaus progressively shift towards right for increasingly values of $V_{DS}$. It may be noted that this results in a 1-DEG formation threshold voltage which increases with increasing drain voltage contrary to the negative $V_T$ roll-off observed due to short channel effects. In addition, the fringe field is electrostatically coupled to the channel from all three sides of the nanowire as the depth of 1-DEG channel is (~3-5 nm) comparable with the width of the nanowire (~10 nm). This is further verified through two-dimensional solution of Schrodinger and Poisson equations in our device geometry (Fig. 1(d)). The threshold voltage increases approximately linearly for larger nanowire width and gate to nanowire separation. The magnitude of threshold voltage can be effectively reduced by using an insulator of high dielectric constant, smaller separation between the nanowire and gate, and a smaller width of the nanowire (see supplementary document). A dielectric encapsulation may be used to further reduce the magnitude $V_T$.

The $I_{DS}$ versus $V_{DS}$ characteristics for the transistor is shown in Fig. 4(a). It may be noted that the spacing of the change in saturation drain current for equal changes in ($V_{GS}-V_T$) is neither parabolic (long channel pinch-off for bulk transistors) nor linear (pinch-off due to velocity saturation). This is primarily due to the significant additional contribution from quantum capacitance in this 1-D system, which itself depends on the $V_{DS}$ and $V_{FG}$. The carrier concentration in the 1-D channel is not proportional to the geometric capacitance as in 3-D and 2-D transistors, which leads to this apparent anomaly. The quantum capacitance also brings in non-linearity for small values of $V_{DS}$, which is also a direct contrast to conventional transistors. It is further interesting to note that the current in the 1-D channel suddenly drops to near zero when $V_{GS}$ changes from –16 to –18 V, which is a signature of a sharp pinch-off of the 1-D channel. To further confirm 1-D operation and fringe field control, devices are also made with long channel width W = 100 nm, which essentially does not provide any quantum confinement in the lateral direction. The typical output characteristics of such a transistor is shown in Fig. 4(b). It may be noted that the characteristics show linear behavior for small values of $V_{DS}$ and the saturation drain current varies almost linearly with $V_{GS}$-$V_T$. The corresponding transfer characteristics and $g_m$ are shown in Fig. 4(c), which do not show any plateau in $I_{DS}$ and $g_m$-oscillations, respectively.

In summary, we have demonstrated a novel method for the fabrication of site-controlled lateral nanowires in AlGaN/GaN based heterostructures. The current flow through the nanowire bears the signature of 1-D transport which is effectively controlled by a fringe gate. The transfer characteristics show plateaus interleaved by sharp changes, which translate to oscillations in the transconductance characteristics due to subband resolved transport controlled by the fringe gate. The experimental observations are coherently explained by taking into account quantum capacitance in addition to the geometric capacitance.

This work is partially supported by DST, Indian Space Research Organization (ISRO) and Advanced Semiconductor Technology (ASemiT, DRDO) lab.

Figure Captions:

Fig 1: (Color online) (a) A schematic of the heterostructure used in our work; (b) a schematic of the fringe gate nanowire transistor; (c) equilibrium band diagram for the AlGaN/GaN nanowire; (d) simulated electric field pattern around the nanowire; (e) an SEM image of a nanowire is shown in the inset. A cross-sectional SEM image of a nanowire; (f) a colored SEM image of the fringe gate nanowire transistor; (g) an array of nanowire transistors; (h) PL of the nanowire sample after blanket etching.

Fig. 2: (Color online) $I_{DS}$-$V_{GS}$ characteristics in (a) linear and (b) log scale; (c) transconductance of the nanowire transistor.

Fig. 3: (Color online) (a) 1-DEG electron density and quantum capacitance of a nanowire for large intersubband separation; (b) quantum capacitance and transconductance characteristics of the nanowire; (c)-(e) shows the band-diagrams under various conditions.

Fig 4: (Color online) (a) $I_{DS}$-$V_{DS}$ characteristics of a nanowire transistor; (b) $I_{DS}$-$V_{DS}$ and (c) $I_{DS}$-$V_{GS}$ and $g_m$ of a control transistor having a large width.

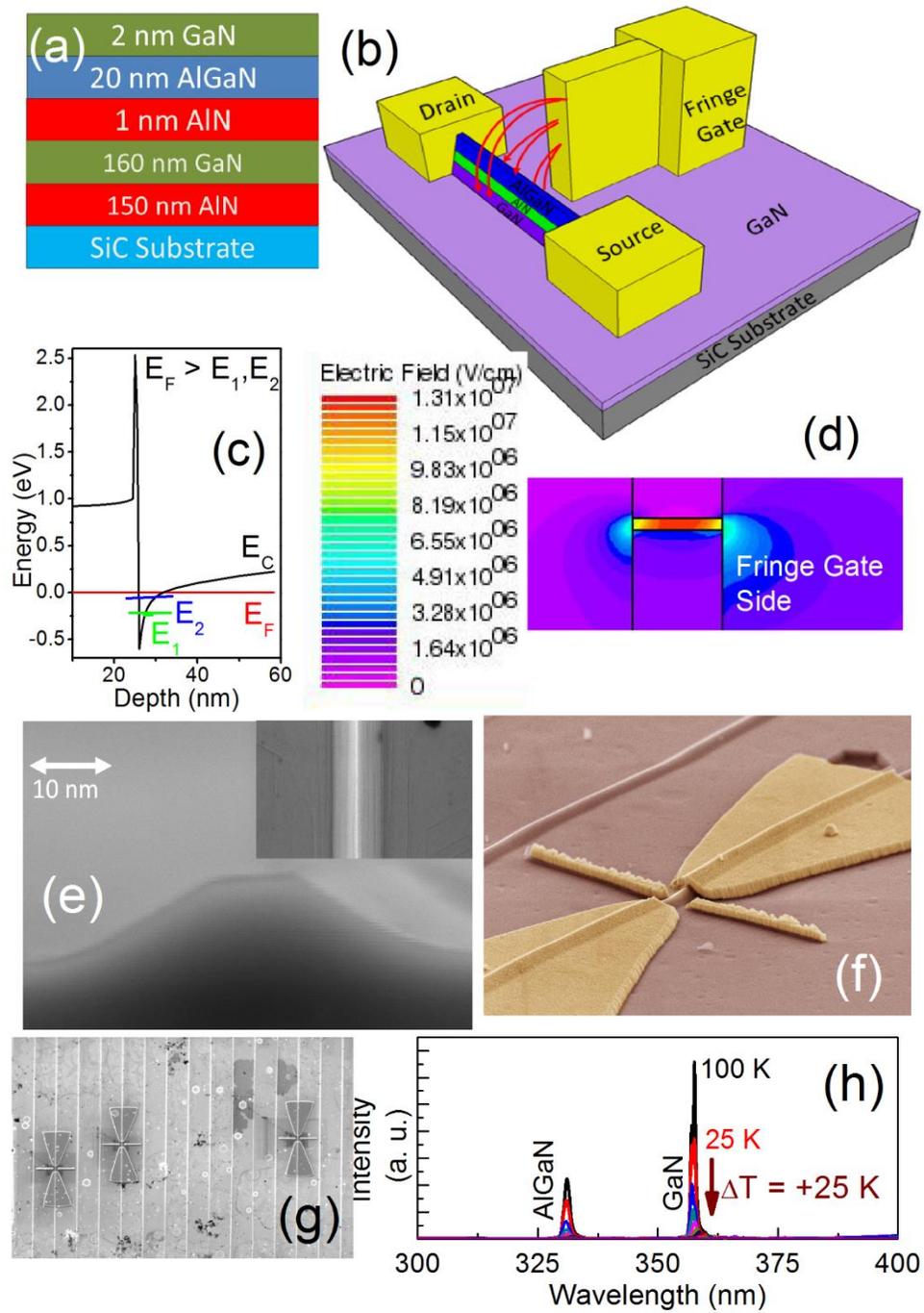

Figure 1 of 4
S *et al.*

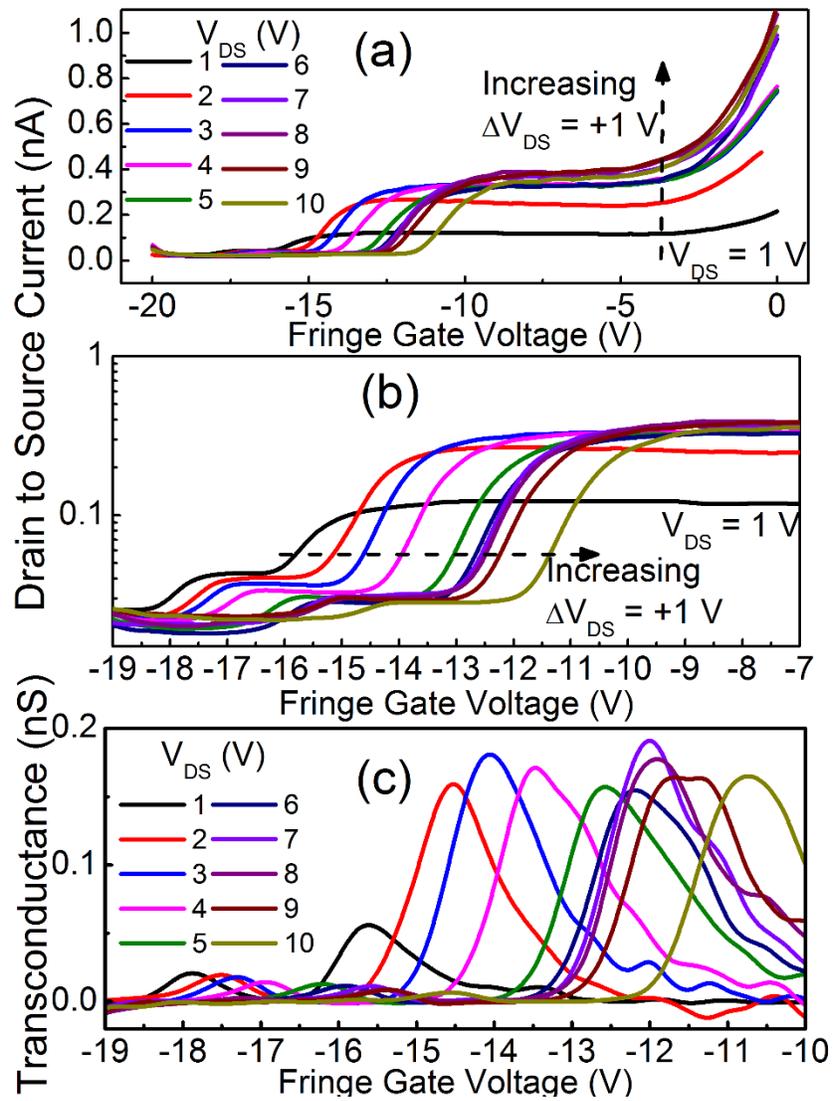

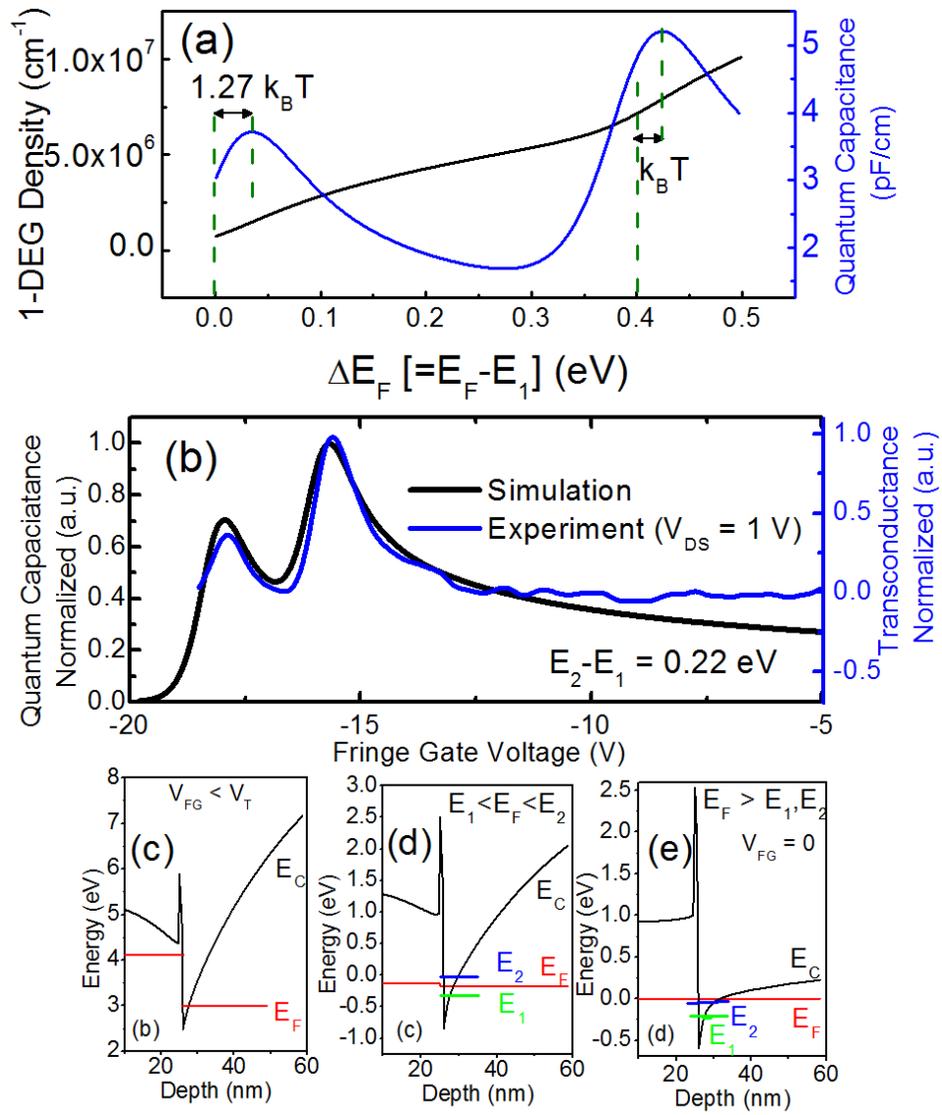

Figure 3 of 4
S *et al.*

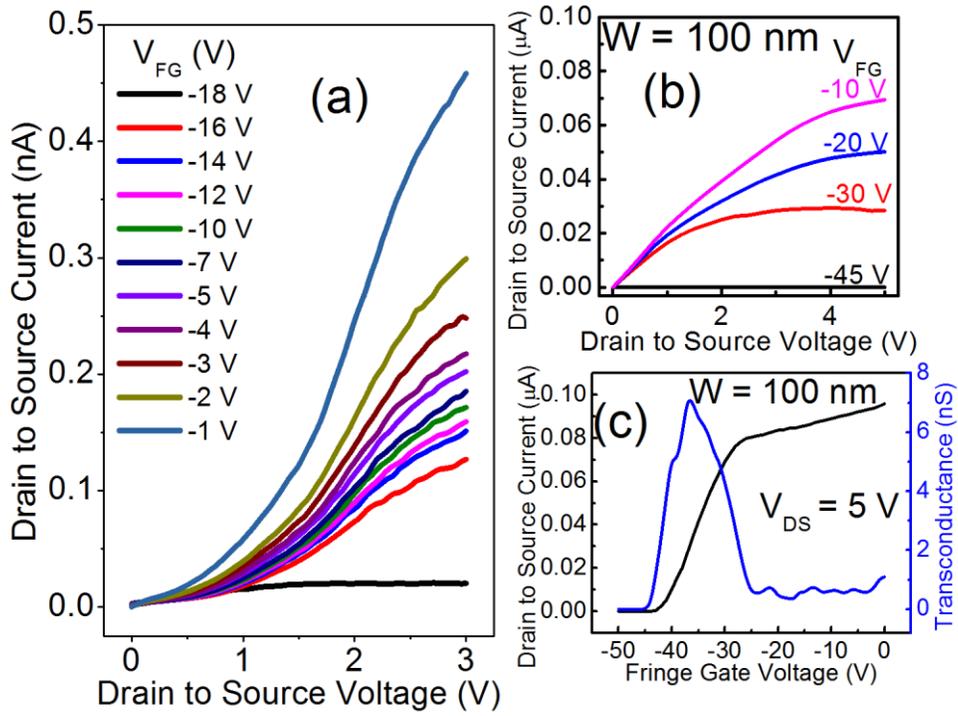

# Additional Information

## A. Nanowire Structural Characteristics

The longer dimension of the nanowire is aligned with the sides of the hexagonal pit. This alignment ensures nanowires of very large aspect ratio with smooth edges. Supplementary Fig. 1(a) shows a typical magnified image of a nanowire in the vicinity of a hexagonal pit. The nanowires are trapezoidal in shape. A magnified image of the nanowire is shown in supplementary Fig. 1(b).

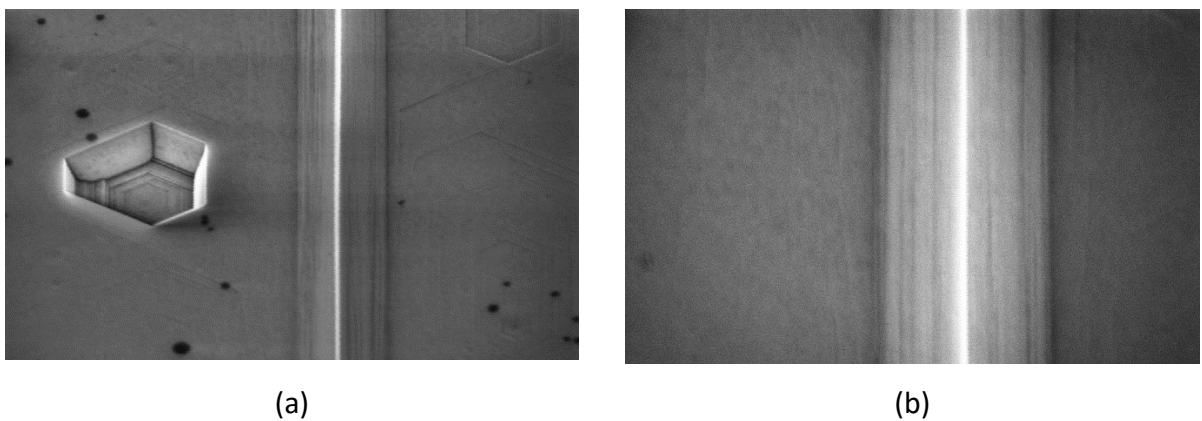

(a) (b)

Supplementary Fig. 1: (a) An SEM image of a nanowire in the vicinity of a hexagon; (b) a high magnification SEM image of a nanowire.

## B. Nanowire Sidewall Orientation

The sidewalls of the nanowires have miller indices of $(10\bar{1}3)$ and $(\bar{1}013)$. The experimentally observed angle between them was found to be 119° which is very close to the theoretically expected value of 116°.

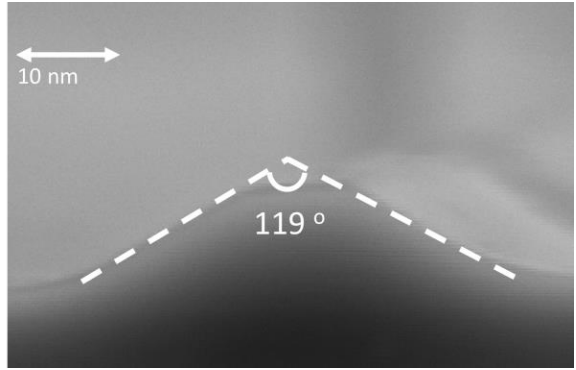

Supplementary Fig. 2: The angle between the sloping sidewalls of the nanowire was found to be 119°.

## C. Device Fabrication

A rectangular region of 1.0 μm × 0.1 μm is delineated by using electron-beam lithography (EBL), which determines the site of the nanowire. The sample is etched in an inductively coupled reactive-ion-etcher (ICP-RIE) using Ar:$Cl_2$ (10:20 sccm) at ICP power 500 W, RF power 60 W and pressure 0.67 Pa for a depth of 100 nm. It is then treated in boiling $H_3PO_4$ for 25 s. The sidewalls of etch much faster (~0.2 μm/min) that the top AlGaN layer (~3 nm/min) which lead to the formation of the nanowires. The miller indices of the exposed side-walls are $(10\bar{1}3)$ and $(\bar{1}013)$. The nanowires are formed at the center of the original rectangular region. The nanowire is aligned with the side of a hexagon which allows several microns long nanowire formation with very small width. Ohmic contacts are formed at the end of the nanowires by electron-beam (e-beam) evaporation of Ti/Al/Ni/Au (30 nm/130 nm/30 nm/100 nm) and rapid thermal annealing at 850 °C for 30 s in $N_2$ environment. Finally, the fringe gate is formed by e-beam evaporation of Ni/Au (30 nm/600 nm). Fringe gate devices with various nanowire dimensions of 5-100 nm are fabricated by this technique where the gate is kept a fixed distance from the nanowire.

## D. Quantum Capacitance of 2-D and 1-D Systems

The quantum capacitance is computed for a GaN based hypothetical 2-D system for a width W = 10 nm and a 1-D system. There is a reduction in the electron density for the 1-DEG is noted. The quantum capacitance in shown in the right y-axis. The quantum capacitance is significantly lower for the 1-D system and hence it plays a more important role for nanowire transistors than in the case of 2-DEG based high electron mobility transistors. The smaller quantum capacitance makes the AlGaN/GaN based nanowire transistors useful to study 1-D quantum transport. The peak of the quantum capacitance occurs at $E_F-E_1 = 1.27\ k_BT$. The position and relative magnitude of the second peak is dependent upon the separation $\Delta E = E_2-E_1$. The distinct peaks may disappear when $\Delta E$ is small.

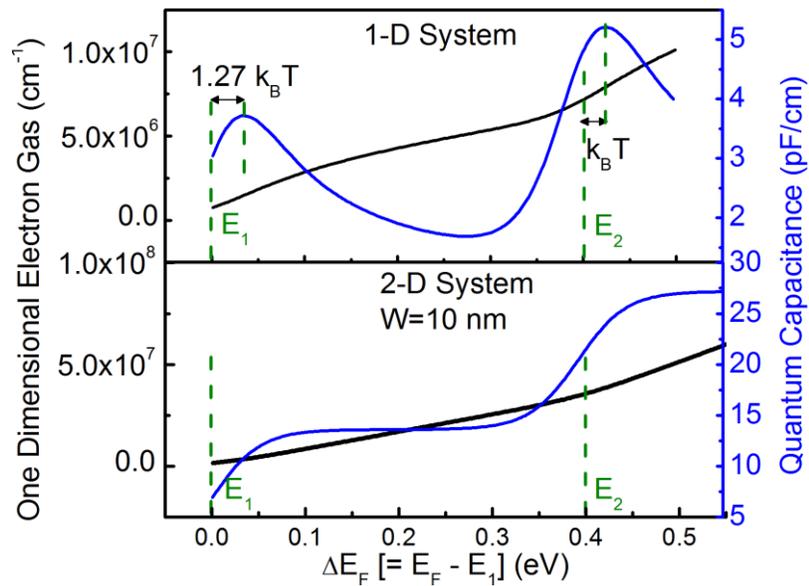

Supplementary Fig. 3: Charge density and quantum capacitance for 1-D and hypothetical 2-D (W = 10 nm) systems. The separation of bound states is arbitrarily kept large.

E. Threshold Voltage

The threshold voltage for 1-DEG formation increases with increasing with $V_{DS}$. Supplementary Fig. 4 shows the approximate linear dependency as observed for the nanowire transistor.

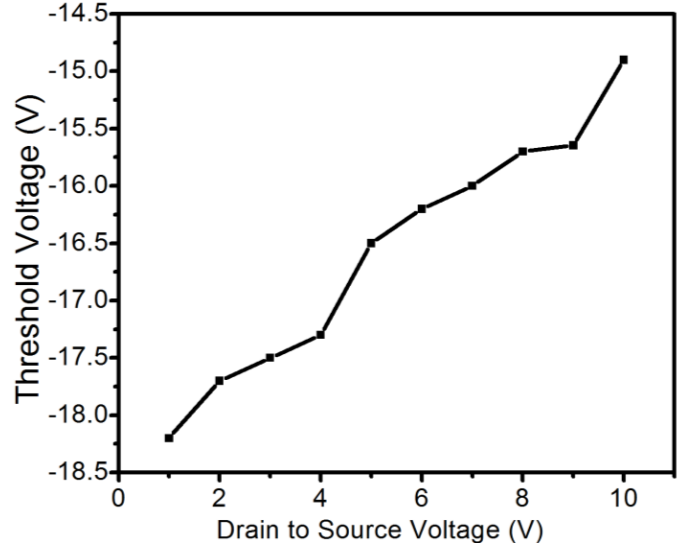

Supplementary Fig. 4: Threshold voltage as a function of $V_{DS}$ for the nanowire transistor.

**F. Transconductance $g_m$ and Quantum Capacitance in the Linear Region for 1-D System**

The current through the channel in the linear region with very small drain to source voltage is given by:

$$I_{DS} = q n_s \mu E$$
$$= q n_s \mu \frac{V_{DS}}{L}$$

where q is the electronic charge, $n_s$ is the 1-DEG density, µ is the mobility, E is the electric field, $V_{DS}$ is the drain to source voltage, and L is the channel length. The transconductance $g_m$ of the nanowire transistors is given by

$$g_m = \frac{dI_{DS}}{dV_{GS}} = C_T \mu \frac{V_{DS}}{L}$$

where $C_T$ is total capacitance. The $C_T$ is given by the expression:

$$C_T = q \frac{dn_s}{dV_{GS}} = q \frac{dn_s}{dE_F} \frac{dE_F}{dV_{GS}}$$

where $E_F$ is the Fermi level. The first differential term in the above expression corresponds to the quantum capacitance ($C_q = q^2 dn_s/dE_F$). The $n_s$ is related to $E_F$ through 1-D density of states and Fermi-Dirac statistics. Combining the above equations lead to Eqn. 1. of the manuscript.

## G. Oscillations in $C_q$

The maxima and minima of the quantum capacitance are determined by solving the following equation numerally:

$$\frac{d^2}{dE_F{}^2} \sum_i \left[ \int_{E_i}^{\infty} \frac{1}{\sqrt{E-E_i}} \frac{1}{1+\exp\left(\frac{E-E_F}{k_B T}\right)} dE \right] = 0$$

where $E_i$ denotes the energy of the $i^{th}$ 1-D subband.

## H. Threshold Voltage

Devices with a fixed dimension of ~10 nm and various gate separations are also fabricated for analyzing the effect of 3-D electrostatics on the device characteristics. Variations of the threshold voltage as a function of wire width and fringe gate distance are shown in Supplementary Fig. 5.

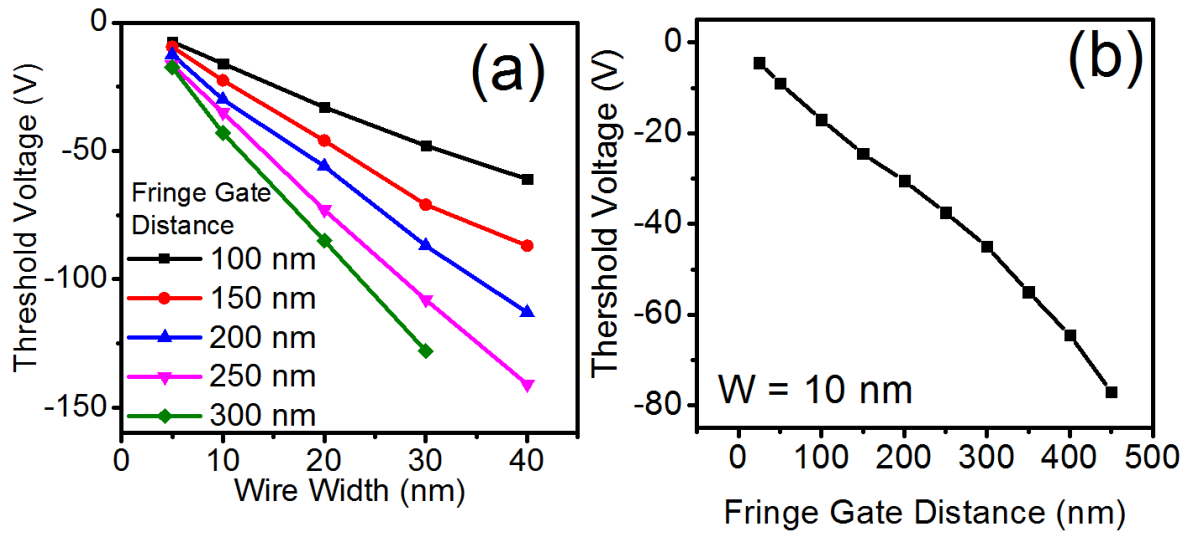

Supplementary Fig. 5: Variation of thershold voltage as a function of (a) wire width and (b) fringe gate distances.